\documentclass[10pt]{article}
\usepackage {latexsym}
\usepackage {amsmath}
\usepackage {amscd}
\usepackage {amssymb}
\usepackage{graphics}
\usepackage{psfrag}

\newcommand{\pb}[2]{\left\{#1, #2\right\}_{\rm{PB}}}
\newcommand{\db}[2]{\left\{#1, #2\right\}_{\rm{D}}}
\newcommand{\Xx}{\mathbb{X}}
\newcommand{\Yy}{\mathbb{Y}}
\newcommand{\column}[2]{ \left( {\begin{array}{*{20}c} {#1}  \\ {#2}  \\ \end{array} } \right) }
\newcommand{\Qq}{\mathcal{Q}}
\newcommand{\dirac}{\not{\partial}}

\usepackage{psfig}

\begin{document}
\thispagestyle{empty}
\rightline{\small EMPG-06-04}
\rightline{\small hep-th/0605114}
\vspace*{2cm}

\begin{center}
{\bf \LARGE Quantum Mechanics of the Doubled Torus}\\
\vspace*{1cm}

{\bf Emily~Hackett-Jones}\footnote{E-mail: {\tt
e.hackett-jones@ed.ac.uk}}
{\bf and George~Moutsopoulos}\footnote{E-mail: {\tt g.moutsopoulos@sms.ed.ac.uk}}

\vspace*{0.3cm} \vspace{0.5cm}
School of Mathematics, University of Edinburgh \\
Mayfield Road, Edinburgh EH9 3JZ\\
Scotland\\

\vspace{2cm}
{\bf ABSTRACT}
\end{center}

We investigate the quantum mechanics of the doubled torus
system, introduced by Hull \cite{Hull} to describe T-folds in a more geometric way. Classically, this
system consists of a world-sheet Lagrangian together with some constraints, which
reduce the number of degrees of freedom to the correct physical
number. We consider this system from the point of view of constrained
Hamiltonian dynamics. In this case the constraints are second class,
and we can quantize on the constrained surface using Dirac
brackets. We perform the quantization for a simple T-fold background
and compare to results for the conventional non-doubled torus system. Finally, we formulate a consistent
supersymmetric version of the doubled torus system, including
supersymmetric constraints.

\vfill
\setcounter{page}{0}
\setcounter{footnote}{0}
\newpage

\section{Introduction}

In recent years there has been considerable interest in defining
string theory on various  ``duality-folds''\cite{Hull, Kumar:1996zx, Hellerman:2002ax, Dabholkar:2002sy,
Kachru:2002sk,Hull:2003kr, Flournoy:2004vn, Hull:2005hk, Gray:2005ea,
Shelton:2005cf, FlournoyWilliams, Dabholkar:2005ve, Lawrence:2006ma,
Hull:2006qs, Walcher}.
These backgrounds differ from ordinary manifolds because field
configurations  on overlapping coordinate patches can be glued
together using duality transformations, as well as the conventional
diffeomorphisms and gauge transformations. Of particular interest are
backgrounds involving the T-duality group of transformations; these
are  known as ``T-folds''~\cite{Hull}. These backgrounds are $n$-torus fibrations
over some base, where the fibre undergoes
monodromy transformations in the T-duality group $O(n,n;\mathbb{Z})$
around certain cycles in the base. T-folds
are therefore fibre bundles with structure group
$O(n,n;\mathbb{Z})$. A key feature of T-folds is that, unlike
manifolds, they do not possess a globally well-defined metric. This is
because T-duality transformations  mix up the metric and B-field
components.  Nevertheless, sensible supergravity compactifications
(typically using the Scherk-Schwarz ansatze~\cite{Scherk:1978ta, Scherk:1979zr}) can be defined on these
backgrounds~\cite{Dabholkar:2002sy,
   Hull:2003kr, Hull:2005hk, Gray:2005ea, Shelton:2005cf, Maharana:1992my, Kaloper:1999yr}.

 In this paper we will be interested in the world-sheet description of
 T-folds. In particular, we will use a framework introduced by Hull
 known as the ``doubled torus'' formalism~\cite{Hull}. Essentially, the idea is to
 double the dimension of the $T^n$ fibre, and consider the
 T-fold as a $2n$-dimensional torus fibration over the same base. Such
 ideas have also been implemented in earlier works, for example in Refs.~\cite{Duff:1989tf, Duff:1990hn}. The
 extra $n$ dimensions are associated to the
 T-dual coordinates, $\tilde{X} = X_L-X_R$.  By enlarging the fibre in
 this way, the monodromy transformations in $O(n,n;\mathbb{Z})$ act linearly.
 Moreover, since $O(n,n;\mathbb{Z})$ is a subgroup of $Gl(2n,
 \mathbb{Z})$, which is the group
 of large diffeomorphisms of $T^{2n}$, T-folds are geometric backgrounds from the doubled torus
 perspective. Physically speaking, one can think of the
 doubled torus as the set of all possible T-duals of a given T-fold~\cite{Hull}.

Now, since the dimension of the fibre has been doubled, one must
impose constraints to halve the number of physical degrees of
freedom in order to make contact with critical string theory. In Ref.~\cite{Hull}
covariant constraints with the right properties are introduced.
Therefore, the doubled torus model consists of a world-sheet
Lagrangian, together with some constraints. These constraints can be
imposed in a number of ways. One way is to solve the constraints and
re-write everything in terms of the physical degrees of freedom. Using
this approach Hull~\cite{Hull} shows that {\it classically}  this leads to the
conventional non-doubled formulation. In particular, by solving the
constraints and using them in the doubled torus equations of motion,
one arrives at equations of motion for a sigma model on the
non-doubled torus, $T^n$.
Furthermore, by
choosing different ``polarizations'' for the physical coordinates one can obtain sigma models related to the original one by T-duality. In particular, the B\"{u}scher rules for the transformation of the metric and
B-field under T-duality can be recovered.

Our approach will be to investigate the doubled torus formalism as a
constrained Hamiltonian system. In particular, we will not solve the
constraints, but rather we will impose them on our Hamiltonian, and
then move to
Dirac brackets in order to quantize on the constrained surface. The
 first aim of this paper is to investigate whether the doubled torus
system is equivalent quantum mechanically to the more conventional
non-doubled torus.

One hope is that the doubled torus formalism might be somewhat
simpler quantum mechanically than the non-doubled torus. In the
conventional formalism it is well known that understanding T-folds
quantum mechanically involves the study of asymmetric orbifolds,
which are non-trivial (see for example, Refs.~\cite{FlournoyWilliams, Walcher, Narain:1986qm, Aoki:2004sm}). However, since T-folds are geometric
backgrounds from the doubled torus perspective, one might expect
that there are no asymmetric orbifolds to deal with.
This does not turn out to be the case. In fact, we recover exactly the same (asymmetric) orbifolds and partition
functions as in the conventional case, even though the steps along the
way are somewhat different.

The second aim of this paper is to find the supersymmetric version of
the doubled torus model. Although there is much work on supersymmetric
sigma models~\cite{Witten:1993yc, AbouZeid:1997cw, Abou-Zeid:1999em, Hori:2000kt, Hori:2002fa}, in this case one has to also consider how to make
the constraints supersymmetric. We will construct a consistent
supersymmetric Lagrangian with suitable constraints.

The plan of this paper is as follows. Firstly, in \S~\ref{review} we review the work of
 Ref.~\cite{Hull}. Then in \S~\ref{constraintsection} we move
to the Hamiltonian formulation and determine the class of constraints
we are dealing with. After establishing that the constraints are
second class we move to Dirac brackets.
 Since the Dirac brackets are very simple we are able to
quantize canonically, without invoking BRST quantization (which would
involve making our constraints first class), or other more complicated
methods. To actually perform the quantization we consider a very simple
T-fold in \S~\ref{orbifold} from the doubled perspective. Our
quantization takes place on the constrained surface, which is a
 surface in phase space. An attractive feature of our analysis is that
 it does not require a choice of polarization to be made.
 We calculate all the quantum mechanical ingredients such as
Virasoro operators, the Hilbert space, the partition function and so
on, and compare to the non-doubled results. In \S~\ref{supersymmetric} we give our results for a supersymmetric version of the
doubled torus formalism, including the supersymmetrized constraint. In
\S~\ref{conclusion} we discuss our results and conclude.

Note added: After our paper first appeared on the archive, a new paper
by C. Hull~\cite{Hull:2006va} appeared which discusses important aspects of the
doubled torus formalism. These include the quantum equivalence to the
usual formulation, arbitrary genus worldsheets and the dilaton. The
method of quantization involves gauging half of the currents, and is
different to the method used here.

\section{Bosonic Theory and Constraint Analysis}\label{review}
We begin by considering the doubled torus system defined by Hull
in Ref.~\cite{Hull}. This is a constrained Lagrangian system, where
the degrees of freedom on the fibre are doubled; constraints
are then imposed to reduce these degrees of freedom to the correct physical
number. We will analyse this setup as a
constrained {\it Hamiltonian} system. This will lead to a natural
quantization in terms of Dirac brackets.

\subsection{Review of doubled torus formulation}
In this section we review the doubled torus construction for
T-folds. The starting point is to consider a sigma model, defined by embedding coordinates
($X^I$, $Y^m$) which map the world-sheet
into the target space. Locally, the target space takes the form
$N\times T^{2n}$, where $T^{2n}$ is the
doubled torus. Globally, however, the target space is a $T^{2n}$ fibre
bundle over
 $N$, with structure group $O(n,n;\mathbb{Z})$. The embedding
coordinates $X^{I}$ are associated to $T^{2n}$, hence we have the
periodicity conditions\footnote{This does not mean that all radii
  in the $T^{2n}$ are equal, but rather the radii will enter in the metric $H_{IJ}$}  $X^I \sim X^I + 2\pi$, where $I=1, \dots, 2n$. The coordinates $Y^m$ are associated to the base,
so $m=0, 1, \dots, 26-n$. Our total number of dimensions is $26+n$,
but $n$ of these will be unphysical.

The data on the
target space consists of a generalised metric, $H_{IJ}$, and source
terms, $J_I$, on $T^{2n}$, together with a metric, $G_{mn}$, and
$B$-field, $B_{mn}$, on the base. The following
Lagrangian can be constructed for this system~\cite{Hull},
\begin{equation}
{\mathcal L}=\frac{1}{2}H_{IJ}(Y)P^I\wedge \star P^J + P^I \wedge
\star J_I(Y) +{\mathcal L}(Y) \label{firstLag}
\end{equation}
where $P^I=dX^I$, and $d$, $\wedge$ and $\star$ are all operations on
the worldsheet. The Lagrangian on the base space, $\mathcal{L}(Y)$,
can be
taken to have the following general form,
\[
{\mathcal L}(Y)=\frac{1}{2}G_{mn} dY^m \wedge \star dY^n +\frac{1}{2} B_{mn} dY^m \wedge dY^n
\]
while the source terms can be expressed in the following natural way, $J_I =A_{In} dY^n + \tilde{A}_{In} \star
dY^n$.
Then the Lagrangian is
\begin{eqnarray}
{\mathcal L}&=&\frac{1}{2}H_{IJ}\partial_a X^I \partial_b X^J \eta^{ab}
 +A_{In}\partial_a X^I \partial_b Y^n \eta^{ab}
 - \tilde{A}_{In}\partial_a X^I \partial_b Y^n
\epsilon^{ab}\nonumber\\
&&+\frac{1}{2}G_{mn}\partial_a Y^m \partial_b Y^n \eta^{ab}
 -\frac{1}{2} B_{mn}\partial_a Y^m \partial_b Y^n \epsilon^{ab} \label{Lagrangian}
\end{eqnarray}
where $\sigma^{a,b} = \tau,\sigma$ are the world-sheet coordinates,
  $\eta = {\rm diag}(+1, -1)$ is the flat world-sheet metric and
  $\epsilon_{01}=+1$. All fields in the above Lagrangian are assumed
  to depend on the base coordinates, $Y^m$, in general.

By varying $X^I$ one obtains the following equation
of motion,
\begin{equation}
d \star (H P + J) = 0 \label{eom}
\end{equation}
This can be written more explicitly as
\[
\eta^{ab}\partial_a (H_{IJ} \partial_b X^J + A_{In} \partial_b Y^n) -
\epsilon^{ab}\partial_m \tilde{A}_{In}\partial_a Y^m \partial_b Y^n =0
\]
Physical solutions of this equation should also satisfy the
following constraint (which is really $n$ constraints), which halves the number of physical degrees of
freedom
\begin{equation}
\star P^I= S^I{}_J P^J + L^{IJ} J_J \label{constraint1}
\end{equation}
where $L^{IJ}$ is a constant $O(n,n)$
invariant metric\footnote{$O(n,n)$
matrices $M$ must therefore satisfy $M^T L M = L$.} and $S^I{}_J
\equiv L^{IK}H_{KJ}$. In the next section we will see precisely why this constraint halves the number of  degrees of freedom. First, however, we see that for the consistency of the constraint
one must have $S^2=1$ and $SL\star J=-LJ$. This restricts the form of
$H_{IJ}$ and implies
\begin{equation}
A_{In} = - H_{IJ}L^{JK}\tilde{A}_{Kn} \label{consistency}
\end{equation}
for the constituents $A, \tilde{A}$ of the source term $J_I$.

An important feature of the doubled torus system is that it is
invariant under $O(n,n;\mathbb{Z})$. Suppose we consider a global $O(n,n)$
transformation, $M$ (which must satisfy $M^T L M = L$), then $X, H, J$ transform as follows,
\begin{eqnarray}
X &\rightarrow&  M X \nonumber\\
H &\rightarrow&  (M^{-1})^T H M^{-1}\nonumber\\
J &\rightarrow& (M^{-1})^T J
\end{eqnarray}
Hence the Lagrangian (\ref{firstLag}) and constraint
(\ref{constraint1}) remain invariant under the continuous group
$O(n,n)$. However, only the
discrete subgroup $O(n,n;\mathbb{Z})$ will leave the lattice for $X^I$
invariant.

To make contact with the conventional formulation of bosonic strings on $T^n$,
one must divide the
coordinates on $T^{2n}$ into $n$ physical coordinates, $X^i\in T^n$,
and $n$ dual coordinates, $\tilde{X}_i \in \tilde{T}^n$. This is
referred to as a ``choice of polarization''. In group
theoretic language, we are decomposing $O(n,n)$ into representations of
$GL(n)$. In particular, the $2n$-dimensional representation of
$O(n,n)$ decomposes as $2n \rightarrow n + n'$, where $n$ and $n'$ are
the fundamental and anti-fundamental representations of $GL(n)$. This
decomposition can be implemented in a geometric way by the following
$2n\times 2n$ matrix \cite{Hull},
\[
\Pi = \left( \begin{array}{c}
\Pi^i{}_I\\
\tilde{\Pi}_{{i} I}\end{array}\right)
\]
where upper $i$ indices correspond to the $n$ representation, and
lower ${i}$ indices correspond to $n'$. The physical subspace $T^n$ must
be a null subspace with respect to the constant  $O(n,n)$ metric $L$,
i.e.
\begin{equation}
\Pi^i{}_I \Pi^j{}_J L^{IJ} = \tilde{\Pi}_{{i} I} \tilde{\Pi}_{{j} J} L^{IJ} = 0 \nonumber
\end{equation}
Also,
\[
\Pi^i{}_I \tilde{\Pi}_{i J} + \tilde{\Pi}_{i I} \Pi^i{}_J = L_{IJ}
\]
In terms of the $GL(n)$ basis, $L_{IJ}$ can always be taken to be
\begin{equation}
L_{IJ} = \left( \begin{array}{cc}
0& 1_{n\times n}\\
1_{n\times n} & 0\end{array}\right)
\end{equation}
which gives the natural metric $ds_L^2 = 2 dX^i
d\tilde{X}_i$. Moreover, in a certain gauge \cite{Hull} the metric $H_{IJ}$ can be chosen to take the conventional form for a $O(n,n)/O(n)\times O(n)$
coset metric, namely
\begin{equation}
H = \left( \begin{array}{cc}
G-BG^{-1}B & BG^{-1} \\
-G^{-1}B & G^{-1}
\end{array}\right)
\end{equation}
where $G$ and $B$ are the metric and B-field on
 $T^n$. Notice that if we take $L$ and $H$ as above then
 $S^2 = 1$ automatically, and also ${\rm Tr} S =0$.

The distinguishing feature of T-folds, compared to ordinary manifolds, is that in general no global polarization, $\Pi$, can be chosen, even though it is always possible locally. This is equivalent to the earlier statement about T-folds, namely that they have no globally well defined metric. Hence the above form for $H_{IJ}$ only makes sense locally.

At this point we take a different approach to
Ref.~\cite{Hull}. Instead of solving the constraint
(\ref{constraint1}) for $\tilde{X}$ in terms of the other quantities,
we will consider the doubled torus as a
constrained Hamiltonian system. In particular, we will determine the
class of the constraint we have here, and then use methods from
constrained Hamiltonian dynamics to quantize on the constrained
surface.

\subsection{Hamiltonian and Constraint Analysis}\label{constraintsection}

We can write the total Lagrangian (\ref{Lagrangian}) in a more compact form as
\begin{equation}
{\mathcal L}=\frac{1}{2} g_{\mu\nu}\dot{q}^{\mu}\dot{q}^{\nu} +
\dot{q}^{\mu} j_{\mu} - V[q] \label{compactLagrangian}
\end{equation}
where the indices $\mu=I,n$, we define $q^{I}=X^I$, $q^{n}=Y^n$ and
$\dot{q}^\mu \equiv \partial_\tau q^{\mu}$. The metric is
\begin{equation}
g_{\mu\nu} = \left(\begin{array}{cc}
H_{IJ}& A_{In}\\
A_{Jm}& G_{mn}\end{array}\right)
\end{equation}
The source terms $j_{\mu}$ are given by
$j_I=\tilde{A}_{In}Y'^n$, $j_n=-\tilde{A}_{In}X'^I+B_{nm}Y'^m$, and
the potential is $V[q]=\frac{1}{2}g_{\mu\nu}q'^{\mu}q'^{\nu}$, where
$q'^{\mu} \equiv \partial_{\sigma} q^{\mu}$.

The conjugate momenta are
$\pi_{\mu}=g_{\mu\nu}\dot{q}^{\nu}+j_{\mu}$. More explicitly, the
conjugate momentum
of $X^I$ is
\[\pi_I=\frac{\partial L}{\partial\dot{X}^I}=H_{IJ} \dot{X}^J+A_{In}\dot{Y}^n+\tilde{A}_{In}Y'^n\]
and the conjugate momenta of $Y^n$ is
\[\pi_n=\frac{\partial L}{\partial\dot{Y}^n}=G_{mn}\dot{Y}^m+A_{In}\dot{X}^I-\tilde{A}_{In}X'^I+B_{nm}Y'^m\]
This allows us to calculate the Hamiltonian density, ${\mathcal H}$. We find,
\begin{align}
{\mathcal H}&=
 \pi_{\mu}\dot{q}^{\mu}-{\mathcal L}\nonumber \\&= \frac{1}{2}g^{\mu\nu}(\pi_{\mu}-j_{\mu})(\pi_{\nu}-j_{\nu})+\frac{1}{2}g_{\mu\nu}q'^{\mu}q'^{\nu}
\end{align}
Here we see that the Hamiltonian is only well-defined if $g_{\mu\nu}$
is invertible. Note that $H_{IJ}$ and $G_{mn}$ being invertible do not
guarantee that $g^{\mu\nu}$ exists. However, for our
analysis we will require $g_{\mu\nu}$ to be invertible.

We now discuss the constraint which we want to impose on this
Hamiltonian system. Recall that the constraint (\ref{constraint1}) is
\[
\star P^I= S^I{}_J P^J + L^{IJ} J_J
\]
Writing it in its two components gives
\[\Phi^{-}_1=P_{\tau}-S P_{\sigma}-L J_{\sigma} = 0 \]
\[\Phi^{-}_2=P_{\sigma}-S P_{\tau}-L J_{\tau} = 0\]
where we are omitting $I, J$ indices for brevity. Taking sums and
differences of these two equations, one finds
\[
\frac{1}{2}(1 \pm S)(P_\tau \mp P_\sigma) = \frac{1}{2} L (J_\sigma \mp J_\tau)
\]
Now since $S^2 =1$ and ${\rm Tr} S =0$, this means that $(1 \pm S)/2$
are projectors onto two orthogonal $n$-dimensional subspaces.
Therefore, the
constraint is forcing half of the $X^I$s to be purely left moving, and
half to be purely right moving.

In fact, using $S^2=1$ and $SL\star J=-LJ$, one finds $\Phi^-_2=-S\Phi^-_1$
and so we can take $\Phi^-_1$ as our only primary constraint. Using
the consistency conditions one finds we can rewrite $\Phi^-_1$ as follows,
\[
\Phi^{-I}_1 = H^{IJ} (\pi_J - L_{JK} P^K_{\sigma})
\]
where $\pi_J$ is the conjugate momentum for $X^J$ defined
above. Therefore, our primary constraint can be taken to be in the form
\begin{equation}
\Phi_I^{-}\equiv \pi_I-L_{IJ}X'^J\label{constraint2}
\end{equation}

We now calculate the Poisson bracket of the constraint $\Phi^{-}$ with
various other quantities. This will allow us to determine whether this
constraint is first or second class. Recall that the canonical Poisson brackets are
\[\pb{X^I(\sigma)}{\pi_J(\sigma')}=\delta^I_J\delta(\sigma-\sigma')\]
\[\pb{Y^n(\sigma)}{\pi_m(\sigma')}=\delta^n_m\delta(\sigma-\sigma')\]
We consider the closure and time evolution of the constraint. We find,
\begin{align*}
\pb{\Phi_I^-(\sigma_1)}{\Phi_J^-(\sigma_2)} &= -2L_{IJ}\delta'(\sigma_1-\sigma_2)\\
\pb{\Phi_I^-(\sigma)}{\int_{\sigma'} \mathcal{H}} &= \partial_{\sigma}(-L_{IJ}H^{JK}\Phi^-_K)\simeq 0
\end{align*}
This means our constraint, $\Phi_I^-$, is second class. By imposing it we can safely reduce our theory on the constrained phase
space $\Phi_I^-=0$ leaving no other symmetry or gauge freedom. On that surface the dynamics are described by the
Dirac bracket,
\[
\db{A}{B}=\pb{A}{B}-\int_{\sigma,\sigma'}\pb{A}{\Phi_I^-(\sigma)}G^{IJ}(\sigma,\sigma')\pb{\Phi_J^-(\sigma')}{B}\]
where
\[G^{IJ}(\sigma,\sigma')=\pb{\Phi^{-}_I(\sigma)}{\Phi^{-}_J(\sigma')}^{-1}=-\frac{1}{4} L^{IJ}\left(\epsilon(\sigma-\sigma')-\epsilon(\sigma'-\sigma)\right)\]
and $\epsilon$ is the Heaviside step function.
We find the following
Dirac brackets,
\begin{align*}
\db{X^I(\sigma)}{X^J(\sigma')} &= -\frac{1}{4}L^{IJ}\left(\epsilon(\sigma-\sigma')-\epsilon(\sigma'-\sigma)\right)\\
\db{X^I(\sigma)}{\pi_J(\sigma')} &= \frac{1}{2}\delta^I_J\delta(\sigma-\sigma')\\
\db{\pi_I(\sigma)}{\pi_J(\sigma')} &= \frac{1}{2}L_{IJ}\delta'(\sigma-\sigma')
\end{align*}
Moreover, we can define the rotated coordinates $\Phi^{+}_I=\pi_I+
L_{IJ}X'^J$, and the Dirac brackets of $\Phi_I^{\pm}$ are given by
\begin{align*}
\db{\Phi_I^{-}(\sigma)}{A} &= 0\\
\db{\Phi_I^{-}(\sigma)}{\Phi_J^{+}(\sigma')} &= 0\\
\db{\Phi_I^{+}(\sigma)}{\Phi_J^{+}(\sigma')} &= 2 L_{IJ}\delta'(\sigma-\sigma')
\end{align*}
where $A$ is any quantity.
The
coordinates $\Phi_I^{+}$ can be thought of as tangent to the
constraint surface, $\Phi_I^{-} = 0$. In terms of the coordinates $\Phi_I^{\pm}$, the
 Hamiltonian can be written as
\begin{equation}{\mathcal
    H}=\frac{1}{2}g^{\mu\nu}Z_{\mu}Z_{\nu}-\frac{1}{4}H^{IJ}\Phi^{+}_I\Phi^{+}_J+\frac{1}{4}H^{IJ}\Phi^{-}_I\Phi^{-}_J+\frac{1}{2}G_{mn}Y'^m
    Y'^n \label{Hamiltonian}
\end{equation}
with $Z_I=\Phi^{+}_I-\tilde{A}_{In}Y'^n$,
$Z_m=\pi_m-B_{mn}Y'^n$. Notice that $\Phi^-$ only appears quadratically. This will be important later.

\section{Bosonic
Orbifold}\label{orbifold}
In this section we quantize a simple example of a T-fold from the
doubled, constrained Hamiltonian perspective. In the non-doubled
language the T-fold we are interested in has a $S^1$ fibre over an $S^1$ base, with a T-duality
acting on the fibre as one traverses the base.
First, we describe the
doubled description of this background, including the relevant orbifold; then we discuss how to implement the
Dirac brackets quantum mechanically.

\begin{picture}
(0,0)%
\includegraphics{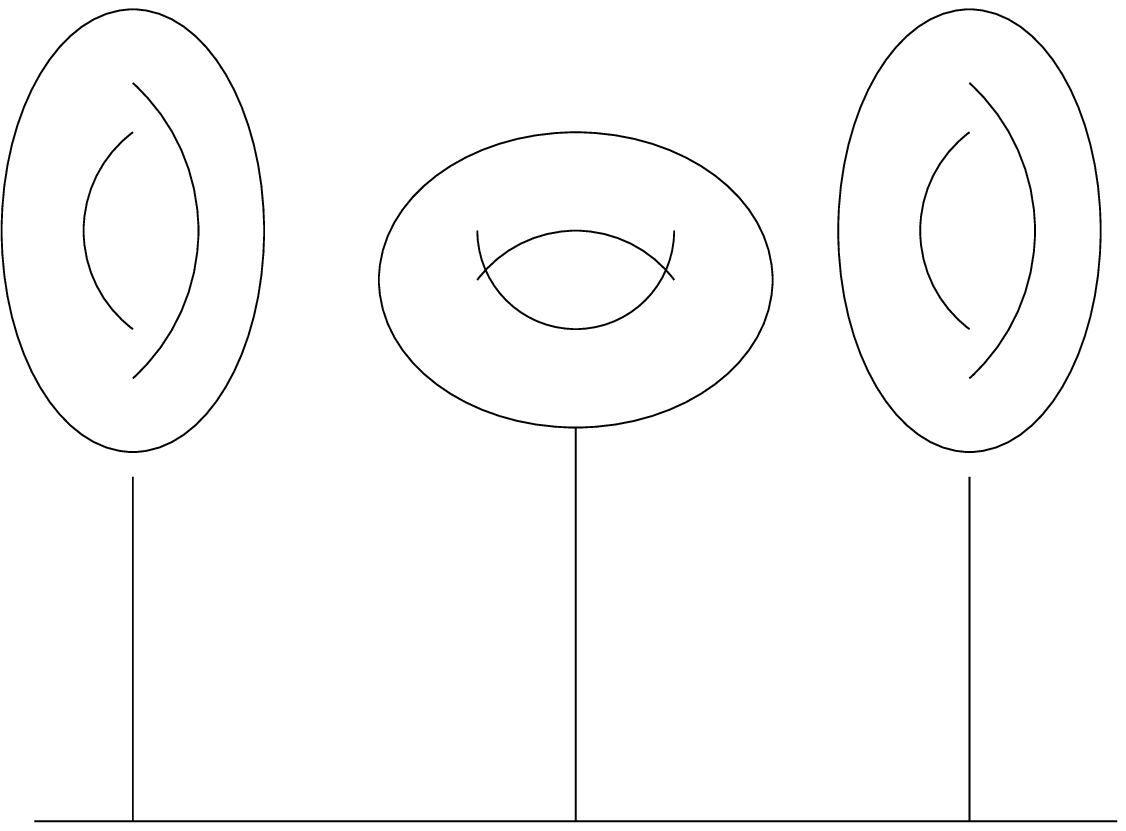}%
\end{picture}%
\setlength{\unitlength}{4144sp}%
\begingroup\makeatletter\ifx\SetFigFont\undefined%
\gdef\SetFigFont#1#2#3#4#5{%
  \reset@font\fontsize{#1}{#2pt}%
  \fontfamily{#3}\fontseries{#4}\fontshape{#5}%
  \selectfont}%
\fi\endgroup%
\begin{picture}(5120,4002)(2993,-3494)
\put(7426,-3436){\makebox(0,0)[lb]{\smash{\SetFigFont{12}{14.4}{\rmdefault}{\mddefault}{\updefault}{$4 \pi R_y$}%
}}}
\put(5626,-3436){\makebox(0,0)[lb]{\smash{\SetFigFont{12}{14.4}{\rmdefault}{\mddefault}{\updefault}{$2 \pi R_y$}%
}}}
\put(3601,-3436){\makebox(0,0)[lb]{\smash{\SetFigFont{12}{14.4}{\rmdefault}{\mddefault}{\updefault}{$0$}%
}}}
\end{picture}

\subsection{The setup}

 In the doubled language our T-fold corresponds to a $T^2$ fibred over $S^1$.
We take the coordinates on the doubled fibre to be $X^1, X^2$, while
$Y$ will be the coordinate on the base $S^1$. As one traverses the
base $S^1$ the fibre undergoes a monodromy transformation, where the
monodromy is the only non-trivial element of
$O(1,1;\mathbb{Z})$. From the non-doubled point of view, this
corresponds to ordinary T-duality on an $S^1$ fibre, i.e. $R
\rightarrow R^{-1}$. As we will see, the monodromy will act
naturally on the doubled coordinates.
Moreover, it constitutes a geometric
transition function for the doubled torus.

So locally our background looks like
\[ N \times S^1 \times T^2 \]
where $N$ is taken to be some flat manifold with coordinates $Y^a$.
For simplicity, we turn all B fields off. That is, we set
\[A_{mI}=\tilde{A}_{mI}=B_{mn}=0\]
and also require no $Y$ dependence
\[\partial_n H_{IJ}=\partial_n G_{mn}=0\]
where $Y^m = (Y^a, Y)$.

We now construct the orbifold using the following identifications.
\begin{eqnarray*}
X^I &\rightarrow&  M^I{}_J X^J \\
Y &\rightarrow&  Y+2 \pi R_y\\
Y^a &\rightarrow&  Y^a
\end{eqnarray*}
where $Y$ is the coordinate on a circle with radius $2R_y$
(i.e. $Y\equiv Y + 4\pi R_y$), so this
corresponds to a half shift around the circle. Therefore, our orbifold
is of order 2. The associated transformation for the metric $H_{IJ}$ is
\[
H \rightarrow  (M^{-1})^T H M^{-1}\]
Using the coset form for the metric $H$, we have
\[
H = \left( {\begin{array}{cc}
   {R^2 } & 0  \\
   0 & {R^{ - 2} }  \\

 \end{array} } \right)
\]
where $R$ is the radius of the original $S^1$ fibre (in the
non-doubled picture). Here the monodromy matrix $M\in
O(1,1; \mathbb{Z})\subset GL(2; \mathbb{Z})$ will be the only
non-trivial possibility, namely $M^I{}_J=L^{IJ}$:
\begin{equation}M=\left( {\begin{array}{cc}
   0 & 1  \\
   1 & 0  \\
 \end{array} } \right)\label{M}
\end{equation}
Therefore, using the transformation rule for $H$, we see that our monodromy corresponds to
\[R\rightarrow  \frac{1}{R}\]
which is what we want.
\subsection{Equations of Motion}

We now consider this particular doubled torus orbifold  from the point of view of the constrained
Hamiltonian system. Recall that we began with phase space
$(X^I,\pi_I,Y^n,\pi_n)$ but on the reduced surface $\Phi^{-}=0$, so the
phase space is $(\Phi_I^{+},Y^n,\pi_n)$. Henceforth we put
$\Phi^{-}_I=0$ and use the symbol $\Phi_I$ for $\Phi^{+}_I$. From
(\ref{Hamiltonian}) the
Hamiltonian is
\[\mathcal{H}=\frac{1}{2}g^{\mu\nu}Z_{\mu}Z_{\nu}-\frac{1}{4}H^{IJ}\Phi_I\Phi_J+\frac{1}{2}G_{mn}Y'^m Y'^n\]
with $Z_I=\Phi_I$, $Z_m=\pi_m$ here. The
non-trivial Dirac brackets are
\begin{align*}
\db{Y^n(\sigma)}{\pi_m(\sigma')} &= \delta^n_m\delta(\sigma-\sigma')\\
 \db{\Phi_I(\sigma)}{\Phi_J(\sigma')} &= 2 L_{IJ}\delta'(\sigma-\sigma')
\end{align*}
The equations of motion $\dot{f}=\db{f}{\int_{\sigma'} H(\sigma')}$ are
\begin{align}
\dot{\Phi_I} &= L_{IJ}H^{JK}\Phi'_K   \label{eq1}\\
d\star d Y &= d\star d Y^a =0 \label{eq2}
\end{align}
Note that the equation of motion for $\Phi_I$ is different from
equations of motion one would obtain from the doubled torus Lagrangian. This is
because we are now considering dynamics on the constrained surface.
We can solve (\ref{eq1}) by diagonalising $L H^{-1} = S^T$ into $\pm
1$ eigenspaces.
\[S^T =L H^{-1}= \left( {\begin{array}{cc}
   0 & {R^2 }  \\
   {R^{ - 2}} & 0 \\
 \end{array} } \right)\]
Then we obtain the following solution for $\Phi_I$,
\[
\Phi_I(\sigma,\tau)=\Phi_{0I}+\Phi_I^{(+1)}(\sigma^+)+\Phi_I^{(-1)}(\sigma^-)\]
where $\Phi_{0I}$ is constant, and $\Phi_I^{(\pm 1)}$ are $\pm 1$
eigenvectors of $S^T$.
The periodicities of
$\Phi_I^{(\pm 1)}(\sigma^{\pm})$ will be determined by the particular boundary
conditions we choose. Note that $\Phi_I$ will not have any linear
terms in $\sigma^{\pm}$. This is because $\Phi_I = \Pi_I + L_{IJ}X'^J$
and both $\Pi_I$ and $X'^J$ are periodic.

The solution for (\ref{eq2}) is
\begin{align}
Y=&Y_R(\sigma^-)+Y_L(\sigma^+)\nonumber\\
Y_R(\sigma^-)=&\frac{1}{2}y_0+p_R\sigma^- +\frac{1}{\sqrt{2}}\sum_{k\neq 0} \frac{i b_k}{k} e^{-i k\sigma^-}\nonumber\\
Y_L(\sigma^+)=&\frac{1}{2}y_0+p_L\sigma^+ +\frac{1}{\sqrt{2}}\sum_{k\neq 0} \frac{i \tilde{b}_k}{k} e^{-i k\sigma^+}\label{Y}
\end{align}
In the next section we will use the boundary conditions for $Y$ to
determine a quantization rule for $p_L$ and $p_R$. Similarly, solving
(\ref{eq2})  for the rest of the coordinates, $Y^a$, one obtains
\begin{equation}
Y^a=y_0^a+p^a\tau + \frac{1}{\sqrt{2}}\sum_{k\neq 0} \frac{i b^a_k}{k} e^{-i k\sigma^-}
+ \frac{1}{\sqrt{2}}\sum_{k\neq 0} \frac{i \tilde{b}^a_k}{k} e^{-i k\sigma^+} \label{Ya}
\end{equation}

For the orbifold we are interested in here, we can distinguish two
twisted sectors. These sectors will give us the boundary conditions we
need to fix the solutions $\Phi_I$ and $Y$ completely. Sector I has
\begin{align*}
\Phi_I(\sigma+2\pi)=&\Phi_I(\sigma)\\
Y(\sigma+2\pi)=&Y(\sigma)+ 4 \pi R_y m&m\in\mathbb{Z}
\end{align*}
Sector II has
\begin{align*}
\Phi_I(\sigma+2\pi)=&M_I{}^J \Phi_J(\sigma)\\
Y(\sigma+2\pi)=&Y(\sigma)+ 2 \pi R_y (2m+1)&m\in\mathbb{Z}
\end{align*}
So our two sectors are distinguished by whether we shift an odd or
even multiple of $2\pi R_y$ around the base circle. We will now consider each of these sectors in turn.

\subsection{Sector I}

We have the boundary conditions
\begin{align*}
\Phi_I(\sigma+2\pi)&= \Phi_I(\sigma)\\
Y(\sigma+2\pi)&= Y(\sigma)+ 4 \pi R_y m&m\in\mathbb{Z}
\end{align*}
From these boundary conditions we see that
$\Phi^{\pm}_I(\sigma^{\pm})$ are periodic functions.
Therefore, the solution for $\Phi_I$ is
\begin{equation}
\Phi_I(\sigma,\tau)= \column{q_1}{q_2} + \column{R}{R^{-1}}\sum_{k\neq 0} \tilde{a}_k e^{-i
k\sigma^+}+\column{R}{-R^{-1}}\sum_{k\neq 0} a_k e^{-i k\sigma^-} \label{Phi}
\end{equation}
where the vectors
\[
e_{\pm}(R) \equiv \left(\begin{array}{c} R\\
 \pm R^{-1}\end{array}\right)
\]
are the $\pm 1$ eigenstates of $S^T$. The constants $q_1, q_2$ are
related to the winding and momentum quantum numbers that would appear in the conventional non-doubled formalism. In appendix~\ref{zeromodes} we show that $q_1, q_2$ obey the quantization condition $q_1 q_2 = 2 mn$, where $m, n\in \mathbb{Z}$.

We now turn to the boundary conditions for
$Y$. The solution for $Y$ is given in (\ref{Y}), and the boundary
conditions give
\begin{align*}
p_L+p_R&\in \frac{1}{2 R_y} n_y &\qquad n_y \in \mathbb{Z}\\
p_L-p_R&\in 2 {R_y}w_y  & \qquad w_y \in \mathbb{Z}
\end{align*}

Now we are ready to quantize $\Phi_I, Y^a, Y$. We will
begin with $Y^m = (Y^a, Y)$, $m=0,\dots 24$ with $Y^{24} \equiv Y$. For our background we have $\pi_n =
\eta_{nm}\dot{Y}^m$, and we want to impose the following bracket as an operator
relation,
\[
\{ Y^m(\sigma, \tau), \pi_n(\sigma', \tau)\} =  \delta^m_n
\delta(\sigma -\sigma')
\]
By replacing $\{,\}\rightarrow -i[,]$ we obtain
the following
commutation relations for the modes associated to $Y^m$,
\begin{equation}
[b_k^m,b_l^n]=
k\delta_{k+l}\eta^{mn}\qquad[\tilde{b}_k^m,\tilde{b}_l^n]=k\delta_{k+l}\eta^{mn}\qquad[b_k^m,\tilde{b}_l^n]=0 \label{Yacommutation}
\end{equation}
Now we consider $\Phi_I$. We have the following Dirac brackets,
\begin{align*}
\db{\Phi_I(\sigma,\tau)}{\Phi_J(\sigma',\tau)}=&2L_{IJ}\delta'(\sigma-\sigma')
\end{align*}
Replacing $\db{}{}\rightarrow -i[,]$ one arrives at the following
\[
[a_k,a_l]= k \delta_{k+l}\qquad [\tilde{a}_k,\tilde{a}_l]= k
\delta_{k+l}\qquad [a_k,\tilde{a}_l]=0
\]
The Hilbert space for this sector, denoted by $H_{(+)}$, will be built
on a vacuum $|0>$ that is invariant under the
monodromy, i.e. $M|0>=|0>$. The states we construct will be
``off-shell'' since we haven't yet imposed physical state
conditions. We decompose the Hilbert space for this sector into
eigenspaces $H_{(+)}^{\pm}$ associated to eigenvalues $\pm 1$ under
the orbifold action, $M$.
Under $M$ we have $\Phi_I
\rightarrow M_I{}^J \Phi_J$, so
using the explicit form for $M$ given in (\ref{M}) this corresponds
to the following action on the eigenvectors which
appear in the decomposition (\ref{Phi}):
\begin{eqnarray*}
e_{\pm}(R) &\longrightarrow& \pm e_{\pm}(R^{-1})
\end{eqnarray*}
together with $q_1\leftrightarrow q_2$.
Therefore, the associated action on the modes is
\[
\tilde{a}_k \mapsto \tilde{a}_k, \qquad a_k \mapsto -a_k
\]
As explained in the recent paper Ref.~\cite{Walcher}, the correct action of
T-duality on states involves a non-trivial phase. This
can be shown by  considering the OPE of two $+1$
T-eigenstates and requiring that no $-1$ eigenstates appear on the
right hand side~\cite{Walcher}. In our doubled language the correct
action of T-duality is
\begin{equation}
T | q_1, q_2> = (-1)^{\frac{q_1 q_2}{2}} |q_2, q_1>
\end{equation}
This phase is essential for modular
invariance of the resulting partition function, as we will see.
Hence the Hilbert space for the non-trivial fibre bundle part of the
space-time splits up into $H_{(+)}=H_{(+)}^{+}\oplus
H_{(+)}^{-}$, where
\begin{align*}
&H_{(+)}^{\pm} =\\
&   \left\{ \prod_{i=1}^{N} a_{-n_i} \prod_{j,k,l}
\tilde{a}_{-m_j} b_{-r_k} \tilde{b}_{-s_l} \left(| q_1,q_2; n_y, w_y>
\pm (-1)^{N+n_y + \frac{q_1 q_2}{2}} |q_2, q_1;n_y, w_y>\right) \right\}
\end{align*}
Note that the factor of
$(-1)^{n_y}$ is  due to the $Y$-shift in our orbifold. See
appendix~\ref{zeromodes} for arguments which lead to the quantization
rule $q_1 q_2 = 2mn$.

Before we move on to sector~II we point out a few interesting
features. Firstly, we only have one set of left-moving
modes and one set of right-moving modes from $\Phi_I(\sigma,
\tau)$.
This
means that
there is no need to make a choice of polarization for our quantum
mechanical states.
This is in contrast to the Lagrangian formulation \cite{Hull}, where classically a polarization must be chosen to
make contact with the non-doubled formulation. Here we do not need to
choose polarization because we have moved to the constrained surface
in phase space.

Secondly, we notice that our orbifold
looks very similar to the interpolating orbifolds
considered in \cite{FlournoyWilliams, Walcher}. This suggests that the doubled
torus formalism is equivalent to the conventional non-doubled formulation of
these backgrounds. However, we must work out the precise details since we are quantizing $\Phi$, not $X$, and so there may be differences in, for example, the
physical state conditions or the partition function.

\subsection{Sector II}

In this sector we have the boundary conditions
\begin{align*}
\Phi_I(\sigma+2\pi)&= M_{I}{}^J\Phi_J(\sigma)\\
Y(\sigma+2\pi) &= Y(\sigma)+ (2m+1) 2\pi R_y&m\in\mathbb{Z}
\end{align*}
The solutions for $Y$ and  $Y^a$ are unchanged from sector~I and are given in
(\ref{Y})-(\ref{Ya}). However, the quantization conditions for $p_L, p_R$ are now
\begin{align*}
p_L+p_R &\in \frac{n_y}{2 R_y} &\qquad n_y \in \mathbb{Z}\\
p_L-p_R &\in  {R_y}(2w_y +1) &\qquad w_y \in \mathbb{Z}
\end{align*}
due to the new boundary conditions on $Y$.
The oscillator algebras for the modes associated to $Y$ and $Y^a$ are
unchanged from sector~I and are given
in (\ref{Yacommutation}). We now turn our
attention to $\Phi_I$. The
solution for $\Phi_I$ can still be written as
\[
\Phi_I = \Phi_{0I} + e_+(R) f(\sigma^+) + e_-(R) g(\sigma^-)
\]
but now the boundary conditions imply that $f$ is periodic while $g$
is anti-periodic. Therefore, $\Phi_I$ can be expanded in modes as
\begin{align*}
\Phi_I(\sigma,\tau)=& \column{q}{q} + e_+(R) \sum_{k\neq 0} \tilde{a}_k e^{-i
k\sigma^+}+e_-(R)\sum_{k\in \mathbb{Z}+\frac{1}{2}} a_k e^{-i k\sigma^-}
\end{align*}
where now the boundary conditions for this sector force the constant
 term $\Phi_{01}=\Phi_{02}=q$. The quantization condition on $q$ is
\begin{equation}
q = \frac{1}{\sqrt{2}}\left(n - \frac{1}{2}\right), \qquad
n\in\mathbb{Z} \label{q}
\end{equation}
This condition is chosen so that level matching in this sector makes
sense \cite{Walcher} (see next section).
We
will prove this is the correct quantization in
Appendix~\ref{zeromodes}.

Again we want to impose the following (Dirac) bracket as an operator
relation for the modes,
\[[\Phi_I(\sigma,\tau),\Phi_J(\sigma',\tau)]=2 i
L_{IJ}\delta'(\sigma-\sigma')\]
Note that $\delta' (\sigma - \sigma')$ cannot simply be periodic
since this will not be compatible with the
  monodromy transformations as $\sigma \rightarrow \sigma + 2\pi$.
To get a correct global statement we should replace the right hand
side of the bracket with the monodromy invariant
 \[ 2 i L_{IJ} \delta'(\sigma - \sigma') = i
\left( {\begin{array}{*{20}c}
   {R^{ 2} \left[\delta'_{2\pi } (\Delta\sigma) - \delta '_{4\pi } (\Delta\sigma)\right]} & {\delta '_{2\pi } (\Delta\sigma) + \delta '_{4\pi } (\Delta\sigma)}  \\
   {\delta '_{2\pi } (\Delta\sigma) + \delta '_{4\pi } (\Delta\sigma)} & {R^{-2} \left[\delta '_{2\pi } (\Delta\sigma) - \delta '_{4\pi } (\Delta\sigma)\right]}  \\
 \end{array} } \right)
\]
where $\Delta \sigma \equiv \sigma-\sigma'$ and $\delta_{2\pi},
\delta_{4\pi}$ are delta functions with period $2\pi$ and $4\pi$ respectively.
Then the commutation relations for the modes are
\begin{equation}
[a_k,a_l]= k \delta_{k+l} \qquad [\tilde{a}_m,\tilde{a}_n]=m
\delta_{m+n} \qquad [a_k,\tilde{a}_m]=0
\end{equation}
where $k,l\in \mathbb{Z}+\frac{1}{2}$ and $ m,n\in\mathbb{Z}$.

We now discuss the (off-shell) Hilbert space of sector II, which
will be denoted by $H_{(-)}$. First note that we need a
twisted vacuum for the right-handed module so that the vacuum flips sign
under the monodromy\footnote{This is evident in radial quantization
  and CFT for states to be well-defined.
}, i.e. $M|0>_{-}=-|0>_{-}$. As in sector I the
action of the monodromy on the modes is
\[
\tilde{a}_k \mapsto \tilde{a}_k, \qquad a_k \mapsto  -a_k
\]
In this
sector there is also a non-trivial phase to take into account, namely
\begin{equation}
T |q > = e^{- \frac{i\pi}{8}}(-1)^{q^2} |q > \label{twistedphase}
\end{equation}
This phase has been proved (in the non-doubled formulation) by Hellerman and Walcher \cite{Walcher} using OPE
relations.

So we have the decomposition of the Hilbert space
$H_{(-)}=H_{(-)}^{+}\oplus H_{(-)}^{-}$, into $\pm 1$ eigenstates under the monodromy, where
\begin{align*}
H_{(-)}^{\pm} = &  \left\{\frac{1 \pm (-1)^{N + n_y + n(n-1)/2}}{2}\prod_{i=1}^{N} a_{-n_i} \prod_{j,k,l}
\tilde{a}_{-m_j} b_{-r_k} \tilde{b}_{-s_l} | q , n_y, w_y >_{-}   \right\}
\end{align*}
and $n\in \mathbb{Z}$ is related to $q$ by (\ref{q}). Here the factor of $(-1)^{n_y}$ comes from the $Y$-shift, as before.

\subsection{Physical State Conditions}

In this section we consider the physical state conditions for the
eigenstates we have constructed above for $H_{(\pm)}$. In particular,
we will investigate the level matching conditions, mass formulae and
ultimately the partition function for this particular
doubled torus setup. Our goal is to show that quantizing the doubled
torus using the constrained Hamiltonian systems method is equivalent
to quantizing the non-doubled torus.

To begin, we will calculate the energy-momentum tensor from the doubled torus
Lagrangian (\ref{compactLagrangian}). As usual, this is defined as
\[T_{ab}=\frac{2}{\sqrt{-h}}\frac{\partial \mathcal{L}}{\partial
  h^{ab}}\Big{|}_{h=\eta}\]
where $h_{ab}$ is a general world-sheet metric. One finds,
\begin{equation}
T_{ab} =  g_{\mu\nu}\partial_a q^{\mu} \partial_b q^{\nu} - \frac{1}{2} \eta_{ab}  \eta^{cd} g_{\mu\nu}\partial_c
q^{\mu}
\partial_d q^{\nu}
\end{equation}
Due to Weyl invariance $T_{00}=T_{11}$, so we need only investigate
$T_{00}$ and $T_{01}$. Written in phase space, for generic
$g_{\mu\nu}, j_\mu$, one finds
\begin{eqnarray*}
T_{00}&=&\mathcal{H}\\
T_{01}&=& \pi_{\mu}{q^{\mu}}'=\frac{1}{4}L^{IJ}\Phi_I^{+}\Phi_J^{+}+\pi_m {Y^m}'
-\frac{1}{4}L^{IJ}\Phi_I^{-}\Phi_J^{-}
\end{eqnarray*}
where $\mathcal{H}$ is the Hamiltonian (\ref{Hamiltonian}). From the above form it is clear that since the
elements $T_{ab}$ form a closed algebra of constraints, they will also form a closed algebra on the constraint
surface $\Phi^{-}_I=0$, since $\Phi^{-}_I$ appears quadratically in both $T_{00}$ and $T_{01}$. The same applies if
we switch from Poisson brackets to Dirac brackets.

We use the above results to calculate the energy-momentum tensor for the model we have been dealing with, where $j_\mu = A_{In} = \tilde{A}_{In} = B_{mn}=0$. We set
$\Phi^-_I = 0$ and denote $\Phi^+_I \equiv \Phi_I$. In terms of the coordinates
$\sigma^{\pm}$, the only non-zero components of
$T$ are $T_{\pm\pm}=\frac{1}{2}(T_{00}\pm T_{01})$, given explicitly by
\begin{eqnarray*}
T_{\pm \pm} &=& \frac{1}{8} \left(H^{IJ} \pm L^{IJ} \right)\Phi_I \Phi_J + \partial_{\pm}Y \partial_{\pm}Y + \eta_{ab}
\partial_{\pm}Y^a \partial_{\pm}Y^b
\end{eqnarray*}
where $\partial_{\pm} = \frac{1}{2}(\partial_0 \pm \partial_1)$. We
now substitute in our mode expansions for $\Phi_I, Y, Y^m$ to obtain
Virasoro operators $L_m, \tilde{L}_m$. We will do this for both
sectors, to obtain physical state conditions for twisted and untwisted
states. We
begin with sector~I.

Substituting in the untwisted expansions for $\Phi_I, Y, Y^a$ into the above
gives the following expressions for the Virasoro operators:
\begin{eqnarray*}
L_m &=& \frac{1}{2\pi} \int_0^{2\pi} e^{im\sigma^-} T_{--} d\sigma^-\\
&=& \frac{1}{2} \sum_{k=-\infty}^{+\infty} (a_{m-k}a_k + b_{m-k}b_k +\eta_{ab} b^a_{m-k}b^b_k)
\end{eqnarray*}
where
\begin{equation}
a_0 \equiv \frac{1}{2} \left( \frac{q_1}{R} - q_2 R \right), \qquad b_0 \equiv
\sqrt{2} p_R, \qquad b_0^a \equiv \frac{p^a}{\sqrt{2}} \label{momenta}
\end{equation}
Similarly,
\begin{eqnarray*}
\tilde{L}_m &=& \frac{1}{2\pi} \int_0^{2\pi} e^{im\sigma^+} T_{++} d\sigma^+\\
&=& \frac{1}{2} \sum_{k=-\infty}^{+\infty} \left(\tilde{a}_{m-k}\tilde{a}_k + \tilde{b}_{m-k}\tilde{b}_k +
\eta_{ab}\tilde{b}^a_{m-k}\tilde{b}^b_k\right)
\end{eqnarray*}
where
\begin{equation}
\tilde{a}_0 \equiv \frac{1}{2}\left(\frac{q_1}{R}  + q_2 R \right), \qquad \tilde{b}_0 \equiv
\sqrt{2}p_L, \qquad \tilde{b}_0^a \equiv \frac{p^a}{\sqrt{2}} \label{tildedefs}
\end{equation}
For the normal ordered zero modes $L_0$ and $\tilde{L}_0$ we have
\begin{eqnarray*}
L_0 &=& \frac{1}{8}\left(\frac{q_1}{R} - q_2 R\right)^2 +  p_R^2 + \frac{1}{4}(p^a)^2
+ \sum_{k=1}^{\infty} (a_{-k}a_k + b_{-k}b_k + b^a_{-k}b^a_k)\\
\tilde{L}_0&=& \frac{1}{8}\left(\frac{q_1}{R} + q_2 R\right)^2 +  p_L^2 + \frac{1}{4}(p^a)^2
+ \sum_{k=1}^{\infty}(\tilde{a}_{-k}\tilde{a}_k + \tilde{b}_{-k}\tilde{b}_k + \tilde{b}^a_{-k}\tilde{b}^a_k)
\end{eqnarray*}
Therefore, the level matching condition is
\begin{equation}
\frac{1}{2} q_1 q_2 + p_L^2 -  p_R^2 + \tilde{N} - N = 0
\end{equation}
Note that the first term will be an integer because we have the quantization condition $q_1q_2 = 2mn$, $m,n\in
\mathbb{Z}$.
The mass spectrum formula is
\begin{equation}
M^2 = 2\left( p_L^2 + p_R^2 + \frac{q_1^2}{4 R^2} +
\frac{q_2^2 R^2}{4}  + N +
\tilde{N} - 2\right)
\end{equation}
where the $-2$ arises as the zero point energy of 24 left-handed and 24 right-handed integer
moded bosonic oscillators, which each contribute $-1/24$.

From the mass formula we see that the state $a_{-1} \tilde{a}_{-1}|k^a>$, which corresponds
to the metric component along the fibre, is indeed massless, as one
would expect. However, it belongs to $H_{(+)}^-$, i.e. it has
eigenvalue $-1$ under the orbifold action. Therefore, this state will
be projected out. This is in
agreement with Refs.~\cite{Dabholkar:2002sy} and \cite{FlournoyWilliams},
where it is explained that when there is a non-trivial monodromy the moduli
must take values which are fixed under the action of the monodromy. In our example $R\rightarrow R^{-1}$, so the
component of the metric with both legs in the fibre has fixed value 1. In
orbifold language this means the corresponding state, $a_{-1} \tilde{a}_{-1}|k^a>$, must be projected
out, which is indeed what we find here.

We now consider the energy-momentum tensor and physical state
conditions for the twisted {sector II}. First note that
\[
T_{--}=\frac{1}{8}\left( \frac{q}{R} - q R \right)^2 + \frac{1}{2}\left( \frac{q}{R} - q R \right) \sum_{k\in
\mathbb{Z}+\frac{1}{2}}a_k e^{-ik\sigma^-}+\dots
\]
That is, $T_{--}$ has both integer and half integer modes; therefore it will be neither periodic nor antiperiodic. $T_{++}$
is periodic and we require $T_{--}$ to be periodic. This is only satisfied if $R=1$. We put $R=1$ from now on.
We then obtain the following $L_m$ and $\tilde{L}_m$,
\begin{eqnarray*}
L_m &=& \frac{1}{2\pi} \int_0^{2\pi} e^{im\sigma^-} T_{--} d\sigma^-\\
&=&  \frac{1}{2} \sum_{k\in\mathbb{Z}+\frac{1}{2}} a_{m-k}a_k +
\frac{1}{2} \sum_{k\in \mathbb{Z}} \left( b_{m-k}b_k  + \eta_{ab} b_{m-k}^a b_k^b \right)
\end{eqnarray*}
where $b_0$ and $b_0^a$ are related to the $Y$-momenta via (\ref{momenta}).
Similarly,
\begin{eqnarray*}
\tilde{L}_m &=& \frac{1}{2\pi} \int_0^{2\pi} e^{im\sigma^+} T_{++} d\sigma^+\\
&=&  \frac{1}{2} \sum_{k=-\infty}^{+\infty}\left( \tilde{a}_{m-k}\tilde{a}_k
+ \tilde{b}_{m-k}\tilde{b}_k + \eta_{ab}\tilde{b}^a_{m-k}\tilde{b}^b_k\right)
\end{eqnarray*}
where $\tilde{a}_0 =q$, and $\tilde{b}_0, \tilde{b}_0^a$ are related to the $Y$-momenta via  (\ref{tildedefs}).
For the normal ordered zero modes, $L_0$ and $\tilde{L}_0$, we have
\begin{eqnarray*}
L_0 &=&  p_R^2 + \frac{1}{4}(p^a)^2 + \sum_{k=\frac{1}{2}}^{\infty} a_{-k}a_k + \sum_{k=1}^{\infty}(b_{-k}b_k + b^a_{-k}b^a_k)\\
\tilde{L}_0&=& \frac{1}{2}q^2 +  p_L^2 + \frac{1}{4}(p^a)^2 + \sum_{k=1}^{\infty}(\tilde{a}_{-k}\tilde{a}_k +
\tilde{b}_{-k}\tilde{b}_k + \tilde{b}^a_{-k}\tilde{b}^a_k)
\end{eqnarray*}
The zero point energy for the right-movers will be $-1$, since we have a contribution of $-1/24$ from each of the 24 periodic bosons. On the left hand side the zero point energy is $-45/48$ since we have 23 periodic bosons contributing $-1/24$ and 1 anti-periodic boson contributing $+1/48$.
So the condition on physical states is
\[(\tilde{L}_0-1)|phys>=(L_0-\frac{45}{48})|phys>=0\]
Hence the level matching condition and mass spectrum formula are given by
\begin{eqnarray}
&&\frac{1}{2}q^2 + p_L^2 -  p_R^2 + \tilde{N} - N -\frac{1}{16} = 0\\
&&M^2 = 2\left( p_L^2 + p_R^2 + \frac{1}{2}q^2 + N + \tilde{N} - (2-\frac{1}{16})\right)
\end{eqnarray}
The term $-1/16$ in the level matching condition looks problematic if the
 formula is written in terms of the original zero mode $q$.
Level matching problems are well known to plague
 asymmetric orbifolds, and generally one must make some kind of fix to
 make the level matching formula sensible. The simplest solution here
 is to quantize $q$ appropriately so that the factor of $-1/16$
 cancels. This happens if we choose $\sqrt{2} q = n - 1/2$, $n\in
 \mathbb{Z}$ \cite{Walcher}. Moreover, in appendix~\ref{zeromodes} we
 show that this quantization rule follows directly from having the
 correct phase (\ref{twistedphase}) for the action of T-duality.  We now move on to investigate
 the partition function for this model. We
 will see that this quantization for $q$ leads to a modular invariant partition function.

\subsection{The partition function}\label{partitionsection}
We now have all the ingredients required to calculate the partition function. We are particularly interested in the partition function for the non-trivial part of the background, namely the fibre bundle over $S^1$. Following Flournoy and Williams \cite{FlournoyWilliams} for the construction of partition functions for interpolating orbifolds, this should be given by
\begin{equation}
Z(\tau) =  \frac{1}{2}\sum_{a,b = 0,1} Z_{(\Phi)}^a{}_b (\tau)
Z_{(Y)}^a{}_b (\tau) \label{partition}
\end{equation}
where $Z^a{}_b$ is the partition trace associated to $b$ insertions, with the trace taken over the Hilbert space $H_a$, i.e.
\[
Z^a{}_b = {\rm Tr}_{H_a} ( g^b q^{L_0} q^{\tilde{L}_0})
\]
Here $g$ is the orbifold action and $q = \exp(2\pi i \tau)$ as
usual. In terms of our previous notation  $H_0 \equiv H_{(+)}$ and
$H_1 \equiv H_{(-)}$.  So the essential point is that we are
multiplying partition traces for the $\Phi$ and $Y$ excitations
together, rather than calculating the full $\Phi$ and $Y$ partition
functions separately and then multiplying the results. This is because
we are dealing with an interpolating orbifold, rather than a simple
product orbifold.

For the $\Phi$ excitations we obtain the following partition traces
from the Hilbert spaces $H_{(\pm)}$ and $L_0, \tilde{L}_0$ found
previously. For sector~I we obtain
\begin{eqnarray}
Z^0{}_0 &=&\frac{1}{|\eta|^2 \sqrt{\tau_2} \epsilon}
\sum_{m,n\in\mathbb{Z}} \exp\left(-\frac{\pi}{\tau_2 \epsilon^2} | m +
n\tau|^2\right)\label{numberone}\\
Z^0{}_1
&=&\left( \frac{2 \eta}{\theta_2}\right)^{1/2}\frac{\overline{\theta_4}(2\tau)}{\overline{\eta}}
\end{eqnarray}
where in both cases we have used the quantization rule $q_1
q_2 = 2mn$, $m,n\in \mathbb{Z}$, which implies
\[
q_1=\sqrt{2}m\epsilon, \qquad q_2 = \frac{\sqrt{2} n}{\epsilon}
\]
for some $\epsilon \in \mathbb{R}$. For $Z^0{}_1$ the only states
which contribute are those with $q_1 =q_2$, which implies $\epsilon =
1$ and $m=n$. For sector~II we obtain
\begin{eqnarray}
Z^1{}_0 &=& \left(\frac{\eta}{\theta_4}\right)^{1/2}\frac{\overline{\theta_2}(\frac{1}{2}\tau)}{\overline{\eta}} \\
Z^1{}_1 &=&
\left(\frac{2
  \eta}{\theta_3}\right)^{1/2}\frac{\overline{\theta_2}(\frac{\tau}{2}; - \frac{1}{4})}{\overline{\eta}} \label{numberfour}
\end{eqnarray}

For completeness we give the partition traces for the $Y$
excitations. These have been given in the following compact form in
Ref.~\cite{FlournoyWilliams},
\[
Z_{(Y)}^a{}_b = \sum_{n, w\in\mathbb{Z}}~\sum_{q=0,1} (-1)^{b q} Z_{2R}\left[ 2n + q\big{|} w+ \frac{a}{2}\right]
\]
where the definition of $Z_{2R}[ \dots | \dots]$ can be found in the
appendix~\ref{theta}.

The partition traces for both the fibre and base directions are
modular covariant, which implies the full partition function
(\ref{partition}) is modular invariant. The modular covariance
properties are
\[
Z(\tau +1)^a{}_b = Z(\tau)^a{}_{b-a}, \qquad Z(-1/\tau)^a{}_b =
Z(\tau)^b{}_{-a}\]
for each $a,b$. To see these conditions are satisfied one
must use some properties of the $\theta$ functions, which are
summarised in Appendix~\ref{theta}.
The modular covariance of the $\Phi$ partition traces relies  both on the
quantization condition for the zero modes and the phase factors, both
which were introduced in Ref.~\cite{Walcher}.
This improves on earlier work
\cite{FlournoyWilliams,Aoki:2004sm} where this orbifold was not found
to be modular covariant.

 So we have
shown that the doubled $S^1$ system, considered as a constrained Hamiltonian
system, is equivalent quantum mechanically to the conventional
non-doubled picture. That is, one obtains the same partition
function. An important point is that we have not needed to make any
choice of physical states. Even though we haven't chosen a polarization it is not
surprising that we obtain the same partition function. This is because T-dual
theories have the same partition function.

\section{The Supersymmetric Doubled Torus}\label{supersymmetric}
An obvious extension to the  doubled torus
formalism is to make the Lagrangian and the associated constraint supersymmetric. This will allow more complicated orbifolds (hopefully
modular invariant, and perhaps realistic) to be considered from the
doubled torus perspective. We have completed the first step, which is simply
to find the supersymmetric doubled torus Lagrangian and
the relevant constraints. However, we leave the problem of constructing
supersymmetric orbifolds from this perspective to future work. Note
that supersymmetric asymmetric orbifolds corresponding to T-folds have
been considered in \cite{FlournoyWilliams, Walcher}, but not from the doubled
formalism/constrained Hamiltonian point of view.

\subsection{Extending the Lagrangian}

We want to make the doubled torus Lagrangian (\ref{Lagrangian}) and the constraints (\ref{constraint1}) supersymmetric. We use the following definitions for
superfields, which are supersymmetric extensions of our $X, Y$:
\[\Xx^I=X^I + \bar{\theta}\psi^I+\frac{1}{2}(\bar{\theta}\theta) F^I\]
\[\Yy^n=Y^n + \bar{\theta}\chi^n+\frac{1}{2}(\bar{\theta}\theta) \phi^n\]
or collectively
\[\Qq^\mu=q^\mu + \bar{\theta}\psi^\mu+\frac{1}{2}(\bar{\theta}\theta) f^\mu\]
Covariant derivatives are defined as follows
\[
D_{\alpha} \Qq^{\mu} = \psi^\mu_{\alpha} + \theta_{\alpha}f^{\mu} -
i(\rho^a \theta)_{\alpha} \partial_a q^{\mu} + \frac{i}{2} \partial_a
(\rho^a \psi^{\mu})_{\alpha} (\bar{\theta} \theta)
\]
Our conventions are given in appendix~\ref{conventions}.
We study the following Lagrangian:
\begin{eqnarray}
{\cal L}&=&\int d^2\theta \left\{ \frac{1}{2} g_{\mu\nu}(\Yy) \bar{D} {\Qq}^\mu D{\Qq}^\nu -\frac{1}{2}
b_{\mu\nu}({\Yy})\bar{D} {\Qq}^\mu (\rho_3) D{\Qq}^\nu \right\}\label{susyLagrangian1}\\
 &=& \int d^2\theta \Big\{\frac{1}{2} H_{IJ}(\Yy)
\bar{D} {\Xx}^I D{\Xx}^J + A_{Im}({\Yy})\bar{D}{\Xx}^I
D{\Yy}^m - \tilde{A}_{Im}({\Yy}) \bar{D} {\Xx}^I (\rho_3) D{\Yy}^m \nonumber\\
&& + \frac{1}{2} G_{mn}(\Yy) \bar{D} {\Yy}^m D{\Yy}^n - \frac{1}{2}
B_{mn}({\Yy})\bar{D} {\Yy}^m (\rho_3) \label{superlag}
D{\Yy}^n\Big\}
\end{eqnarray}
where $\rho_3 = \rho^0\rho^1 = \sigma^3$, the third Pauli
matrix\footnote{Interestingly, $\rho^3\cdot V=-\star V$, where $V = V_a \rho^a$,
that is the volume element acts like the Hodge dual.} and $b_{\mu\nu}$
has non-zero components $b_{Im} = - b_{mI} = \tilde{A}_{Im}$ and
$b_{mn} = B_{mn}$.
 Note that all the spinor indices in the above equations are contracted. We
integrate just the fermionic part, using
$\int d^2\theta (\bar{\theta}\theta)=1$,  to obtain a
supersymmetric Lagrangian. This gives the correct bosonic Lagrangian
upon truncation, i.e. we obtain the original bosonic doubled torus Lagrangian (\ref{Lagrangian}).

In more detail, we expand each superfield term in its constituents. For example,
the first term in (\ref{superlag}) is expanded as follows,
\begin{align*}
H_{IJ}(\Yy)\bar{D}_\alpha\Xx^ID_\alpha\Xx^J=&H_{IJ}(\Yy)\bar{\psi}^I\psi^J+2
H_{IJ}(\Yy)\bar{\psi}^I\theta
F^J \\
&-2 i H_{IJ}(\Yy)(\bar{\psi}^I\rho^a \theta)\partial_a X^J\\
&+H_{IJ}(\Yy)\left(\eta^{ab}\partial_a X^I\partial_b X^J + i \bar{\psi}^I\rho^a\partial_a\psi^J+F^I F^J \right
)\bar{\theta}\theta
\end{align*}
where
\begin{align*}
H_{IJ}(\Yy)=H_{IJ}(Y)+\partial_n
H_{IJ}(Y)\bar{\theta}\chi^n+&\frac{1}{2}\partial_n
H_{IJ}(Y)\bar{\theta}\theta \phi^n\\
&+\frac{1}{2}\partial_m\partial_n H_{IJ}(Y)(\bar{\theta}\chi^m)(\bar{\theta}\chi^n)
\end{align*}
The only terms that contribute to the Lagrangian are those which are coefficients of $\bar{\theta}\theta$ in the expansion.
Expanding everything in this way and integrating we arrive at the following
supersymmetric Lagrangian,
\begin{eqnarray}
{\cal L}&=&\frac{1}{2}g_{\mu\nu}\partial_a q^\mu\partial_b q^\nu\eta^{ab} - \frac{1}{2}b_{\mu\nu}\partial_a
q^\mu\partial_b q^\nu \epsilon^{ab}\nonumber\\
&&+ \frac{1}{2}g_{\mu\nu} i \bar{\psi}^\mu \dirac\psi^{\nu} - \frac{i}{2}b_{\mu\nu}\bar\psi{^\mu}
\rho^3\dirac\psi{^\nu}\nonumber\\
&&+\frac{1}{2}g_{\rho\sigma , \nu}i \bar\psi^\rho \rho^a\psi^\nu\partial_a q^\sigma
-\frac{1}{2}b_{\nu\rho,\mu}i\bar\psi{^\nu}\rho^3\rho^a\psi^\mu\partial_a q^{\rho}\nonumber\\
&&+\frac{1}{2}g_{\mu\nu}f^\mu f^\nu +
\left(-\frac{1}{2}\Gamma^\mu_{\rho\nu}\bar\psi^\rho\psi^\nu-\frac{1}{4}H^{\mu}{}_{\rho\nu}\bar\psi^\rho \rho^3
\psi^\nu\right)g_{\mu\kappa}f^\kappa\nonumber\\
&&-\frac{1}{8}g_{\rho\sigma,\mu\nu}\bar\psi^{\mu}\psi^{\nu}\bar\psi^{\rho}\psi^{\sigma}
+\frac{1}{8}b_{\rho\sigma,\mu\nu}\bar\psi^\rho\rho^3\psi^\sigma\bar\psi^\mu\psi^\nu
\end{eqnarray}
Substituting for $f^{\mu}$ and after some algebra we obtain the
following Lagrangian with auxiliary fields solved,
\begin{eqnarray}
{\cal L}&=&\frac{1}{2}g_{\mu\nu}\partial_a q^\mu\partial_b q^\nu\eta^{ab} - \frac{1}{2}b_{\mu\nu}\partial_a
q^\mu\partial_b q^\nu \epsilon^{ab}\nonumber\\
&&+ \frac{1}{2}g_{\mu\nu} i \bar{\psi}^\mu \not{\nabla}^{+}\psi^{\nu} +
\frac{1}{4}R^{{-}}_{\mu\nu\rho\sigma}\psi^{\mu}_{+}\psi^{\nu}_{+}\psi^{\rho}_{-}\psi^{\sigma}_{-} \label{fullsusylag}
\end{eqnarray}
where $\nabla^{\pm}_{\mu}V^{\nu}=\nabla_{\mu}V^{\nu}\mp \frac{1}{2}H_{\mu}{}^{\nu}{}_{\rho}V^{\rho}$ and
$R^{-}{}^{\mu}{}_{\nu\rho\sigma}=\left [
  \nabla^{-}_{\rho},\nabla^{-}_{\sigma} \right ] ^{\mu}{}_{\nu}$. The
operator $\not{\nabla}^{\pm} = \partial_a q^{\mu} \rho^a
\nabla_\mu^{\pm}$, i.e. the pull-back of $\nabla^{\pm}$ to the
world-sheet. One
could, of course, now expand the above Lagrangian in terms of the original data
$H_{IJ}$, $A_{Im}$, $\tilde{A}_{Im}$, $B_{mn}$. We will not do this here as
the expanded form will not be needed in the following.

\subsection{Supersymmetric Constraints}

We now turn to the constraint. The constraint of the bosonic theory (\ref{constraint1}) can be written equivalently
as
\begin{equation}\label{constraint3}
\dot{X}-L\tilde{A}\dot{Y}=S(X'-L\tilde{A}Y')
\end{equation}
an equation that halves the independent vectors $\{dX^I\}\in X^\star
T(T^{2n})$ on the doubled torus, where $X^\star$ is the pull-back of the
map $X:\Sigma \to T^{2n}$. The
fermions in the supersymmetric sigma model are sections of the $X^\star T(T^{2n})\otimes \sqrt{K}$ bundle and it
is natural to halve the independence of them too. Furthermore, the constraints obtained should be supersymmetric.
We find the following constraint sufficient,
\begin{equation}\label{fermconstraint1}
D_\alpha\Xx^I-L^{IJ}\tilde{A}_{Jn}(\Yy)D_\alpha\Yy^n= - S^I{}_J(\Yy)\rho^3_{\alpha\beta}\left(D_\beta
\Xx^J-L^{JK}\tilde{A}_{Kn}(\Yy)D_\beta\Yy^n\right)
\end{equation}
We also require the same consistency condition (\ref{consistency}) in
its functional form unchanged, i.e.
\[
A_{In}(\Yy) = - H_{IJ}(\Yy) L^{JK} \tilde{A}_{Kn}(\Yy)
\]
The
constraint in (\ref{fermconstraint1}) reduces to (\ref{constraint3})
upon setting fermions and auxiliary fields to zero.

Now we consider
the constraint (\ref{fermconstraint1}) with all fields turned on, at
each order in $\theta$. Firstly, the constant
term reads
\begin{equation}
\psi^I-L^{IJ}\tilde{A}_{Jn}(Y)\chi^n=-S^I{}_J(Y)\rho^3\psi^J +
H^{IJ}\tilde{A}_{Jn}\rho^3 \chi^n \label{thetazero}
\end{equation}
This halves independence of the fermions $\psi^I$ using an endomorphism of the target tangent vector
bundle.
A nice way of writing this is to split the fermions in their chiral
parts. Then the above constraint becomes
\begin{eqnarray}
(1+S)\psi_+ &=&  (1 + S)L\tilde{A}\chi_+\nonumber\\
(1-S)\psi_- &=& (1 - S)L \tilde{A}\chi_-
\end{eqnarray}
These constraints seem very natural as $\frac{1}{2}(1\pm S)$ are projectors. Therefore, half of the $\psi^I$s are
constrained to be given in terms of the $\chi^m$s. From the linear terms in $\theta$ we obtain the following constraints:
\begin{eqnarray}
\dot{X} - S X'- L \tilde{A}\dot{Y}+ H^{-1} \tilde{A}
Y'
&=&  - \frac{i}{2} S \bar\chi^n \rho^1 \partial_n \left( L
\tilde{A}  \chi -  S \rho^3 \psi +
H^{-1} \tilde{A} \rho^3 \chi\right) \nonumber\\
f - L \tilde{A} \phi &=& - \frac{1}{2} \bar\chi^n \partial_n \left( L
\tilde{A}  \chi -  S \rho^3 \psi +
H^{-1} \tilde{A} \rho^3 \chi\right) \nonumber
\end{eqnarray}
The first equation is clearly the initial bosonic constraint
(\ref{constraint3}) on the left hand
side, generalized by the addition of some fermionic terms on the right hand
side. In phase space it can be written in exactly the same way as the
original constraint, namely
\[
\pi_I - L_{IJ}X'^J =0
\]
where $\pi_I$ is the canonical momentum associated to $X^I$ derived
from the supersymmetric Lagrangian (\ref{fullsusylag}). The second
equation above is automatically satisfied when the auxiliary fields
are put on-shell.

 We now turn to the quadratic $\theta$ term of the constraint. In particular, we show how this
is automatically satisfied if the constant and linear terms
are imposed and conserved on shell (i.e. the time derivatives of these
constraints are also satisfied). First note that we can collect the equation of motion for $\Xx^I$ in
supersymmetric form
 from (\ref{susyLagrangian1}) as
\begin{equation}\label{susyeom1}
\bar D_\alpha\left(g_{I\mu}(\Yy)D_\alpha \Qq^\mu - b_{I\mu}(\Yy)\rho^3_{\alpha\beta}D_\beta\Qq^\mu\right)=0
\end{equation}
or
\[\bar D_\alpha\left(H_{IJ}(\Yy)D_\alpha \Xx^J + A_{In}(\Yy)D_\alpha \Yy^n-\tilde{A}_{In}(\Yy)\rho^3_{\alpha\beta}D_\beta\Yy^n
\right)=0\]
Similarly the constraint (\ref{fermconstraint1}) can be written as
\begin{equation}\label{fermconstraint2}
H_{IJ}(\Yy)D_\alpha \Xx^J + A_{In}(\Yy)D_\alpha
\Yy^n-\tilde{A}_{In}(\Yy)\rho^3_{\alpha\beta}D_\beta\Yy^n+L_{IJ}\rho^3_{\alpha\beta}D_\beta\Xx^J=C_{I \alpha}=0
\end{equation}
Note that $\bar D_\alpha \rho^3_{\alpha\beta}D_{\beta}=0$ as a
consequence of the supersymmetry algebra. Therefore, the constraint implies the equations of motion for $\Xx$,
in complete analogy with the bosonic constraint implying the equation
of motion for $X^I$ \cite{Hull}. That is, schematically we have
\[ C_\alpha^I=0 \Rightarrow \bar D_\alpha C_\alpha^I = 0 \Leftrightarrow \rm{eom}(\Xx)\]
By writing the constraint expansion as
\[ C_\alpha^I=C_\alpha^{I(0)}+\bar\theta_\beta
C_{\alpha\beta}^{I(1)}+\frac{1}{2} (\bar\theta \theta) C_\alpha^{I(2)}\]
we can show how
\begin{eqnarray*}
C_\alpha^{I(0)}=0&\textrm{on shell}&\\
C_{\alpha\beta}^{I(1)}=0&\textrm{on shell}&\Longrightarrow C_\alpha^{I(2)}=0\\
\bar D_\alpha C^I_{\alpha}=0&\textrm{eom for }\Xx^I&
\end{eqnarray*}
Thus our supersymmetric constraint (\ref{fermconstraint1}) makes
sense. That is, it halves the fermionic and bosonic degrees of
freedom, without imposing extra unphysical constraints. The two
constraints arising from (\ref{fermconstraint1}) are thus
\begin{eqnarray}
\pi_I - L_{IJ} X'^J &=&0\\
(1+\rho^3 S)_I{}^J g_{J\mu}\psi^\mu &=&0
\end{eqnarray}
where the first equation is our original bosonic constraint plus
corrections.

\section{Conclusion}\label{conclusion}

In this paper we have shown that by applying methods from constrained
   Hamiltonian systems one finds that the doubled torus system is equivalent
   quantum mechanically to the non-doubled system, at least for the
   simple example we have worked out here. Previously, this equivalence
   was only established classically, and these methods had not been
   applied.

The doubled torus system proposed by Hull
   \cite{Hull} is a constrained Lagrangian system, and the natural
   formalism for understanding these systems is the methods of constrained
   Hamiltonian systems, where the dynamics is considered on the constrained
   surface. Therefore, our work is the natural extension of
   Ref.~\cite{Hull} where the Lagrangian formalism was used. By moving to phase
   space, and defining a Poisson structure, we find that we do not need
   to choose a polarization for our new variables $\Phi^+$, and we
   construct a polarization invariant Hilbert space. Making use of the
   results of Ref.~\cite{Walcher} for the action of T-duality on states,
   we find that our Hilbert space leads to a modular invariant
   partition function, which is exactly the same as that of the
   non-doubled theory. This is not surprising since T-dual theories
   should
   have the same Hilbert space and partition fucntions, and the
   doubled torus is, in some sense, the set of all T-duals of a given T-fold.

Note that although we have not needed to choose a
polarization,
if we wanted to interpret our constrained Hamiltonian as a sigma
model without constraints, this would involve choosing a polarization
for $\Phi^+$. In particular, one would need to choose which of the
$\Phi^+$ variables are the momenta.

The zero mode quantization is very interesting. In particular, we
   show that knowing the phase \cite{Walcher} in the action of T-duality leads to the
   correct zero mode quantization. Our construction for proving this
   quantization is an orbifold
   one, as opposed to a more general Wilson line theory such as those proposed in
   Ref.~\cite{Walcher}.

The final part of our paper deals with constructing a consistent
   supersymmetric extension to the doubled torus formalism. This
   involves making the constraint supersymmetric, and then checking
   that the superfield constraint does not impose too many
   restrictions
   on the constituent fields, which would be
   unphysical. Surprisingly, the constraints turn out to be very
   simple, both in the superfield language, and when expanded out as
   coefficients of $\theta$. Our final result is that we have $n$
   bosonic constraints, which contain the original
   constraint plus fermionic corrections, and $n$ new fermionic constraints.

The doubled torus system is a tractable example of a constrained
   Hamiltonian system because its
   Dirac brackets are very simple, allowing us to implement Dirac bracket
   quantization, at least for the simple flat background we have considered.
For curved backgrounds this is generally not possible and one must use a more complicated
   method of quantization, such as that used in
   Ref.~\cite{Hull:2006va}.
In the supersymmetric case we find that everything is very similar to the
   bosonic case, and all of the constraints are second class.
It would be interesting to investigate
   the quantization of the supersymmetric doubled torus and consider associated
   asymmetric
   orbifolds. Note that although we have only considered a very simple
   example of a T-fold, we expect other examples to follow through in
   the same vein, and to also display  quantum
   mechanical  equivalence between the doubled and non-doubled
   formulations.

\section*{Acknowledgements}
We would like to thank James Gray and Harry Braden for useful comments
on a draft version of this paper.
We are extremely grateful to Jos\'{e} Figueroa-O'Farrill for
suggesting this project and for many very helpful
discussions. E~H-J is supported by an EPSRC Postdoctoral Fellowship.

\appendix
\begin{section}{Conventions}\label{conventions}

Our worldsheet metric has signature $(+,-)$. For the Clifford algebra
we define $\{\rho^a,\rho^b\}=2\eta^{ab}$, where $\eta$ is the flat
metric. Whenever needed we will use the representation
\[
\rho^0= \left( {\begin{array}{*{20}c}
   {0} & {-i}  \\
   {i} & {0} \\
 \end{array} } \right)
\]
\[
\rho^1= \left( {\begin{array}{*{20}c}
   {0} & {i}  \\
   {i} & {0} \\
 \end{array} } \right)
\]
In $1+1$ dimensions one has the choice of Dirac, Majorana, Weyl or
Majorana-Weyl spinors. We choose to work with real Majorana
spinors.

Since we are considering $\mathcal{N} =1$ supersymmetry on the
worldsheet, our superfields will involve one Majorana spinor
parameter $\theta_\alpha$, Grassmann odd in nature. The supercharges
are defined as follows
\[Q_\alpha=\frac{\partial}{\partial\bar{\theta}^\alpha}+i(\rho^a\theta)_\alpha \partial_a\]
\[\bar{Q}_\alpha=(Q^*\rho^0)_\alpha=-\frac{\partial}{\partial {\theta}^\alpha}-i(\bar{\theta}\rho^a)_\alpha \partial_a\]
\[\{Q_\alpha,Q_\beta\}=-2 i (\rho^a\rho^0)_{\alpha\beta}\partial_a\]
where $\bar\theta_{\alpha} = \theta_{\beta} \rho^0_{\beta\alpha}$ as
usual. We introduce the super-derivatives
\[D_\alpha=\frac{\partial}{\partial\bar{\theta}^\alpha}-i(\rho^a\theta)_\alpha \partial_a\]
\[\bar{D}_\alpha=(D^*\rho^0)_\alpha=-\frac{\partial}{\partial {\theta}^\alpha}+i(\bar{\theta}\rho^a)_\alpha \partial_a\]
\[\{D_\alpha,D_\beta\}=2 i (\rho^a\rho^0)_{\alpha\beta}\partial_a\]
For these we use the fact that $\overline{(\rho^\alpha\theta)}_\alpha=(\bar{\theta}\rho^a)_\alpha$ and
$\overline{(\frac{\partial}{\partial\theta^\alpha})}=-\frac{\partial}{\partial\bar{\theta}^\alpha}$.
Note that  the super-derivatives anti-commute with the charges,
\[\{Q_\alpha,D_\beta\}=0\]

Our superfields $\Xx$, $\Yy$ are supersymmetric extensions of our $X$, $Y$:
\[\Xx^I=X^I + \bar{\theta}\psi^I+\frac{1}{2}\bar{\theta}\theta f^I\]
\[\Yy^n=Y^n + \bar{\theta}\chi^n+\frac{1}{2}\bar{\theta}\theta \phi^n\]
or collectively
\[\Qq^\mu=q^\mu + \bar{\theta}\psi^\mu+\frac{1}{2}\bar{\theta}\theta f^\mu\]
The covariant derivative of $\Xx$ is given by
\[D_\alpha\Xx^I=\Psi_\alpha^I+\theta_\alpha F^I-i(\rho^a\theta)_\alpha\partial_\alpha
X^I+\frac{i}{2}\partial_a(\rho^a\psi^I)_\alpha\bar{\theta}\theta\]
where we have used the Fierz identity
$\theta_\alpha\bar{\theta}_\beta=-\frac{1}{2}\delta_{\alpha\beta}\bar{\theta}\theta$,
which implies the useful relation
$\bar{\theta}\epsilon_1 \bar{\theta}\epsilon_2=-\frac{1}{2}\bar{\epsilon}_2\epsilon_1 \bar{\theta}\theta$.

\end{section}
\begin{section}{Quantization of the zero modes}\label{zeromodes}
In this section we describe how to obtain the quantization of the zero modes of $\Phi_I$.

First, let's recall the simple case of a quantum point particle on a circle $S^1$, considered as an orbifold
$\mathbb{R}/\mathbb{Z}$. The Hilbert space on $\mathbb{R}$ is made up of momentum states $|p>_{p\in\mathbb{R}}$.
Calling the generator of translations $t:x\rightarrow x+2\pi$, we have
that $t|p>=\exp(i2\pi p)|p>$. The invariant Hilbert space consists of the
projected states
\[\sum_n t^n |p>=\sum_n \exp(i n 2\pi p)|p>=\delta(2\pi p) |p> \]
which implies that $p=0\mod 1$.
If for some reason the momentum was initially quantized in
even integers, on the circle the momentum can be further fractionated to take any integer value. Furthermore, we
want the operator $\exp(i x)$ to be realised on the Hilbert space and this will require all integer values of
momentum to be taken into account.

Now let's turn to sector~I of our model. The constraint (\ref{constraint1}) halves the physical
degrees of momentum, winding and oscillator modes. Because it is a differential constraint, the number of zero
modes of $X^I$ will not be halved. Therefore, we must put in an extra constraint on $X^I_0$, so that we have the
correct number of degrees of freedom of a string theory. The natural constraint to implement is
\[
\Pi^i{}_I X^I_0 = X^i_0, \qquad \tilde{\Pi}_{\underline{i} I} X^I_0 = 0
\]
where $\Pi^i{}_I$ and $\tilde{\Pi}_{\underline{i}I}$ are the projectors
 discussed in our \S~\ref{review}. The indices
$i$ correspond to the physical polarization, and $\underline{i}$ to the unphysical polarization. From
\S~\ref{constraintsection} we have that $X$ and $\Phi$ obey the following Dirac bracket,
\[
\db{X^I(\sigma)}{\Phi_J(\sigma')} = \delta^I_J\delta(\sigma - \sigma')
\]
Hence, once we quantize, we can extract the following commutator
\[
[ X_0^i, \Phi_{0 j}] = \delta^i_j
\]
where $\Phi_{0j}$ is the ``physical'' component. Therefore, $\Phi_{0
  i}$ can be thought of as the conjugate momenta to $X_0^i$. Hence
$\Phi_{0 i}\in \mathbb{Z}$, just as in the case of a quantum point particle on a circle.

For the other polarization, $\Phi_{0\underline{i}}$, we can use the fact that $L^{IJ}\Phi_J \sim 2 X'^I$ (up to
additions of $\Phi^-$ which we have set to zero). Therefore $\Phi_{0\underline{i}}$ obtains the quantization
from the winding modes, and we have $\Phi_{0\underline{i}}\in 2\mathbb{Z}$.

In matrix form, these conditions can be written concisely as
\[
\Pi \Phi_0 = m, \qquad \tilde{\Pi} \Phi_0 = 2n
\]
where we are now thinking of $\Phi$ as a column vector, and $m,n \in \mathbb{Z}$. Then using the relation
\[
(\Pi)^T \tilde{\Pi} + (\tilde{\Pi})^T \Pi = L
\]
we arrive at the covariant quantization condition
\begin{equation}\label{quantzerountw}
\Phi^T_0 L \Phi_0 = 4mn
\end{equation}

We will now show an alternative derivation of this quantization for $\Phi_0$ in sector~I of our $T^2\times \mathbb{R} \times N
/ \mathbb{Z}$ (plus constraint) model. Then we will use the same method to
derive the quantization rule in sector~II. The generator of
$\mathbb{Z}$ will be our orbifold transformation, $g$.
The generator $g$ acts like
$M$ on the fibre and translates by $2\pi R_y$ on the base circle. We write the zero modes as
\[\Phi_0=\column{q_1}{q_2}\]
We have the following action of $g$ on the Hilbert space
\[g|q_1,q_2,n_y>=\exp\left(i\pi( n_y + \frac{q_1q_2}{2})\right) |q_2,q_1,n_y>\]
The factor of $\exp(i\pi n_y)$ is the usual phase coming from the translation on the circle base. The phase
$\exp(i \pi q_1 q_2/2)$ is known to be the right T-duality realisation for closure of OPEs in
sector~I (see eg. \cite{Walcher}). At this stage we don't restrict the
quantization of $q_1$,$q_2$.
After projection with $\sum_n g^n$, the existence of
invariant states requires
\[ \pi\left(\frac{q_1q_2}{2} + n_y \right)= 0 \mod 2\pi \] or for generic $n_y$:
\[ q_1q_2 = 2 m n,\qquad m,n\in\mathbb{Z}\]
This is precisely the quantization condition (\ref{quantzerountw}).

We finally turn to sector~II. We use the results of \cite{Walcher}. In their paper they solve issues
like modular invariance and level matching for asymmetric orbifolds. Our case is what they call ``tame'' and our
starting point is the phase of the T-duality. We write the zero mode as
\[\Phi_0=\column{q}{q}\]
Our generator acts as
\[ g |q,n_y> = \exp\left(i\pi( q^2 - \frac{1}{8}+ n_y) \right)|q,n_y>\]
Our construction is an orbifold one and we can show modular invariance, level matching and quantization of zero
modes by adopting the above phase. The invariant Hilbert space requires (for generic $n_y$):
\[q^2-\frac{1}{8}=0\mod 1\]
The simplest choice with even spacing of the modes $q$ is then
\begin{equation}
q=\frac{1}{\sqrt{2}}\left(n-\frac{1}{2}\right)
\end{equation}
where $n \in \mathbb{Z}$.
\end{section}

\begin{section}{Properties of $\theta$ functions}\label{theta}
The $\theta$ functions we use are given by
\begin{eqnarray}
\theta_2(\tau; z)&=& \sum_{n\in\mathbb{Z}}
q^{\frac{1}{2}(n-\frac{1}{2})^2}e^{i\pi (2n-1) z}\nonumber\\
\theta_3(\tau; z)&=& \sum_{n\in\mathbb{Z}} q^{\frac{1}{2}n^2} e^{i2\pi n z}\nonumber\\
\theta_4(\tau; z)&=& \sum_{n\in\mathbb{Z}} (-1)^n q^{\frac{1}{2}n^2} e^{i2\pi n z}
\end{eqnarray}
where $q = \exp(2\pi i \tau)$ as usual. Usually we will take $z=0$,
and we denote $\theta_i(\tau; 0) \equiv \theta_i$.
The $\theta$ functions can also be written as infinite products as follows,
\begin{eqnarray}
\theta_2(\tau; z)&=& 2 \eta q^{\frac{1}{12}} \cos(\pi z)
\prod_{n=1}^{\infty} (1 - 2 q^n \cos(2\pi z) + q^{2n})\nonumber\\
\theta_3(\tau; z)&=& \eta q^{-\frac{1}{24}} \prod_{n=1}^{\infty} \left( 1 +
2 q^{n - \frac{1}{2}}\cos(2\pi z) + q^{2n -1}\right)\nonumber\\
\theta_4(\tau ; z)&=& \eta q^{-\frac{1}{24}} \prod_{n=1}^{\infty} \left( 1
- 2 q^{n - \frac{1}{2}}\cos(2\pi z) + q^{2n -1}\right)
\end{eqnarray}
where
\begin{equation}
\eta(\tau) = q^{\frac{1}{24}}\prod_{n=1}^{\infty} ( 1- q^n)
\end{equation}
The following modular transformation properties will be useful,
\begin{eqnarray}
\eta(\tau+1) &=&  e^{\frac{i\pi}{12}}\eta(\tau)\nonumber\\
\theta_2(\tau +1; z) &=& e^{\frac{i\pi}{4}}\theta_2(\tau;
z)\nonumber\\
\theta_3(\tau +1;z) &=& \theta_4(\tau;z)\nonumber\\
\theta_4(\tau +1;z) &=& \theta_3(\tau;z)
\end{eqnarray}
as well as
\begin{eqnarray}
\eta\left(-\frac{1}{\tau}\right) &=& ( - i\tau)^{\frac{1}{2}} \eta(\tau)\nonumber\\
\theta_2\left( - \frac{1}{\tau} ;\frac{z}{\tau}\right) &=& ( - i
\tau)^{\frac{1}{2}} e^{\frac{i \pi z^2}{\tau}}\theta_4(\tau; z)\nonumber \\
 \theta_3\left( - \frac{1}{\tau} ;\frac{z}{\tau}\right) &=& ( - i
\tau)^{\frac{1}{2}} e^{\frac{i \pi z^2}{\tau}}\theta_3(\tau;
z)\nonumber\\
\theta_4\left( - \frac{1}{\tau} ;\frac{z}{\tau}\right) &=& ( - i
\tau)^{\frac{1}{2}} e^{\frac{i \pi z^2}{\tau}}\theta_2(\tau; z)
\end{eqnarray}
These properties are what is required to show that (\ref{numberone})-(\ref{numberfour})
satisfy the correct modular covariance properties.

For the $Y$ partition traces we need the following expression \cite{FlournoyWilliams} for $Z_{2R}[\dots |\dots]$,
\begin{align}
&Z_{2R} \left[ 2n + q\big{|} w + \frac{a}{2}\right] =\nonumber\\
&\frac{1}{|\eta(\tau)|^2} \exp\left[ - \pi\tau_2 \left( \frac{(2n +
    q)^2}{4R^2} + 4R^2 (w + \frac{a}{2})^2 \right) + 2\pi i \tau_1 (2n +q)(w + \frac{a}{2})\right]
\end{align}
\end{section}

\end{document}